# Right Thoughts and Right Action: How to Make Agile Teamwork Effective[a]

*By Torgeir Dingsøyr, Diane Strode, and Yngve Lindsjørn*

Teamwork is critical in many industrial sectors. When creating complex software solutions, most companies and public institutions organize work in cross-functional teams and follow the principles of agile development. This approach to knowledge-intensive work seeks to empower team members, ensures that the most competent people make decisions, and manages uncertainty by allowing members to learn and adapt as the work progresses.

Agile methods offer much guidance on teamwork. The Agile Manifesto principles recommend self-organized teams and face-to-face conversations which, according to the principles, is the most efficient and effective method of conveying information. "A great development team," according to a white paper from Scrum.org, "trusts each other" and "pursues technical excellence." [1]

Advice is abundant. For example, Google's re:Work model offers advice to development teams in the form of five key factors for successful teams, including "psychological safety," "structure and clarity," and teamwork that the team members consider meaningful.[2] There is also general advice from years of studies of teamwork and from empirical studies on agile development teams. However, there has yet to be a model that draws together the knowledge from all of these sources and specifically focuses on the effectiveness of agile teamwork.

To fill this gap, we have developed an Agile Teamwork Effectiveness Model (ATEM).[3] Our model is based on a review of empirical studies on agile development teams, general studies of effective teams and teamwork, and practitioner advice. We also incorporated findings from our own two case studies and 22 focus groups. Though primarily intended for collocated agile software development teams, the increasing adoption of agile methods outside IT departments may make the model valuable for other agile workplaces. Why Do We Need a Team Effectiveness Model?

Team effectiveness refers to how team members interact to accomplish their project's goals, while delivering quality work within budget and schedule. Ineffective teamwork is detrimental—it can reduce job satisfaction, interfere with team learning, generate knowledge and skill silos, and generally impede progress.

Teamwork effectiveness models are based on accumulated empirical observations and reasoned arguments, and identify and describe key factors necessary for effective teamwork. Our model, tailored for agile practitioners, offers insights into effective agile teamwork and explains how certain agile practices support it.

The ATEM builds on the "Big Five"[4] model of teamwork effectiveness. It consists of three coordinating mechanisms that facilitate and support five teamwork components critical for team effectiveness (see Figure 1). The ATEM includes observable behaviors that practitioners can use to evaluate teamwork effectiveness (see Table 1 and Table 2) and, if necessary, make informed decisions to improve it.

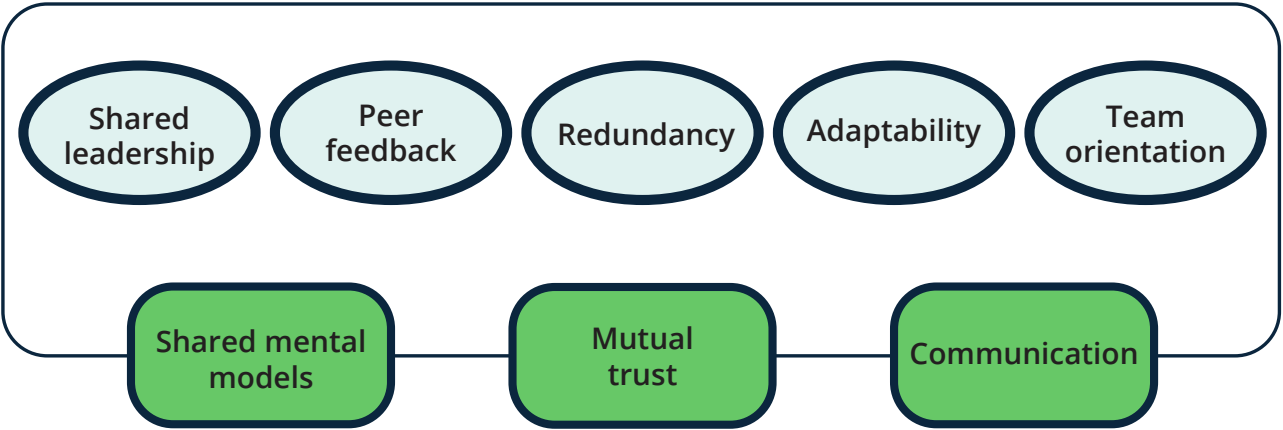

Figure 1. The Agile Team Effectiveness Model (ATEM). Coordination mechanisms are in green, teamwork components are in light blue.

## Coordinating Teamwork

The ATEM coordinating mechanisms—shared mental models, mutual trust, and communication—facilitate and support each other and the five components. For example, a team needs a shared mental model of the work to be done before assisting a team member struggling with workload issues (redundancy); a team needs mutual trust when offering peer feedback to avoid hurt feelings; and a team needs communication to develop a shared mental model and mutual trust. Communication is also vital for each of the five components of teamwork.

| Teamwork coordinating mechanism | Behavioral markers |
| --- | --- |
| Shared mental models<br><br>"An organizing knowledge structure of the relationships among the task the team is engaged in and how the team members will interact." | Anticipating and predicting each other's needs<br><br>Sharing a common understanding of: goals, tasks, the work process, the product, individual skills and expertise |

| Mutual trust | Information sharing |
|---|---|
| "Shared belief that team members will perform their roles and protect the interests of their teammates." | Willingness to admit mistakes and accept feedback<br><br>Supportive team social climate |
| Communication | The team follows up on the progress of tasks |
| "The exchange of information between a sender and a receiver irrespective of the medium." | Visualize project information<br><br>Facilitate informal communication |

Table 1. Coordination mechanisms with behavioral markers in the ATEM. Revised definitions from the "Big Five" model and revised behavioral markers from ATEM.

*Shared mental model* refers to the common understanding that develops among team members over time as they become familiar with each other and the situation. Elements of a shared mental model might be knowledge of the skills and expertise held by team members and knowledge about the situation, the product, the development environment, the process to follow, and the way the team prefers to interact.

An effective shared mental model thus helps coordinate a team, enables smooth working, and ensures fewer interruptions, since team members will spend less time learning from each other by observing or asking and answering questions, and more time on completing tasks.

To evaluate a team's shared mental model, check whether team members can anticipate and predict each other's needs accurately. Do all the team members show knowledge of, or have the opportunity to find out, the goals, the current tasks, and the work process? Are they familiar enough with the product—the stories, design, architecture, and codebase? Are they aware of each other's skills and expertise? When gaps in knowledge are found, focus on practices to fill those gaps: keep teams together over time and integrate newcomers quickly. Hold effective daily stand-up meetings and retrospectives, and promote social activities such as shared lunches. Use pair programming and task boards. Creating an environment where team members can readily express their knowledge, issues, and concerns can promote a common understanding.

*Mutual trust* refers to the shared belief that team members will perform their work roles and protect the interests of their teammates.[4] It is critical for a team's psychological safety[5] (as in the re:Work model), which refers to a climate where people are comfortable being and expressing themselves.

Mutual trust is especially important when teams are self-organizing and rely on empowered members. A lack of trust prompts more formal information sharing within the team. To support mutual trust, the team climate must enable its members to admit mistakes and accept feedback. An unsupportive climate can trigger a more individual orientation, more formal communication, and a more hierarchical project structure.

How can a lack of mutual trust be identified? A poor social climate is one where several conflicts are brewing and team members actively avoid each other. Issues remain unresolved and keep reappearing. A lack of mutual trust can occur when information transfer is formal and not open and when the team members are unwilling to admit mistakes and accept feedback.

*Communication* refers to the exchange of information within the whole team and between individual team members. It is essential in any type of teamwork. A lack of open communication may hinder the sharing of the knowledge and experience necessary for staying abreast of task progress, for offering feedback and support, for discussing issues and identifying solutions, and for understanding the requirements and architectures and translating them into software.

Informal communication is the most effective way for agile teams to share and discuss progress of the tasks so team members are often collocated to stimulate informal and open communication. This makes it easier to visualize information using project boards.

A lack of communication can be identified by checking whether a team has open and informal communication. For a collocated team, study, for example, the physical environment: Does the entire team work in a shared open plan office? Are there team rooms and open spaces that enable the team easy access to each other? Does the team share information openly using physical or virtual boards to visualize task progress, allocation, and dependencies? Does the team hold regular and unscheduled meetings to openly discuss issues and solutions?

# Components of Effective Agile Teamwork

After establishing the coordinating mechanisms, a team must then establish the five key components for effective teamwork: shared leadership, peer feedback, redundancy, adaptability, and team orientation as presented in Table 2. According to our focus group material, development teams are not sufficiently aware of peer feedback and redundancy.

| Teamwork component | Behavioral markers |
|---|---|
| Shared leadership<br><br>"The ability of the team to direct and coordinate their activities, assess team performance, assign tasks, develop team knowledge, skills, and abilities, motivate | The agile team:<br><br>• facilitates team problem-solving<br><br>• determines performance expectations and acceptable interaction patterns |

| Teamwork component | Behavioral markers |
|---|---|
| one another, plan and organize, and establish a positive atmosphere." | - synchronizes and combines individual team member contributions using agile practices combined with automated tools<br>- seeks and evaluates information that affects team functioning<br>- uses agile values and methodologies to determine team member roles<br>- uses agile values and methodologies to determine the frequency and type of preparatory meetings and feedback sessions<br><br>A servant leader facilitates a boundary-spanning function<br><br>Agile team practices provide a planning function |
| Peer feedback<br><br>"The ability to develop common understandings of the team environment and based on those understandings to give accurate peer feedback to team members" | Identifying mistakes and lapses in other team members' actions<br><br>Regular feedback regarding team members' actions to facilitate self-correction |
| Redundancy<br><br>"The ability to anticipate other team members' needs through accurate knowledge about their responsibilities. This includes the ability to shift workload among members to achieve balance during high periods of workload or pressure" | Recognition by potential backup providers that there is a workload distribution problem in their team<br><br>Shifting of work responsibilities to underutilized team members<br><br>Completion of the whole task or parts of tasks by other team members |
| Adaptability<br><br>"The ability to adjust strategies based on information gathered from the | Identifying cues that a change has occurred, assign meaning to that change, and develop a new plan to deal with the change |

| Teamwork component | Behavioral markers |
|---|---|
| environment through the use of backup behavior and reallocation of intra-team resources. Altering a course of action or team repertoire in response to changing conditions (internal or external)" | Identifying opportunities for improvement and innovation for habitual or routine practices<br><br>Remain vigilant to changes in the internal and external environment of the team |
| Team orientation<br><br>"The propensity to take others' behavior into account during group interaction and the belief in the importance of team goal's over individual members' goals"[4] | Taking into account alternative solutions provided by teammates and appraising that input to determine what is most correct<br><br>Increased task involvement, information sharing, strategising, and participatory goal setting<br><br>The team sticks together and remains united |

Table 2. Teamwork components with behavioral markers in the ATEM. Revised definitions from the "Big Five" model and revised behavioral markers from ATEM.

Shared leadership occurs when a team organizes itself to achieve goals. In agile teams, shared leadership can be evenly distributed, making everyone a leader, or rotated among team members based on expertise. Agile teams can also adopt a servant leadership style. In this approach, the servant leader initially empowers the team, then steps back. When the team has to interact with parties outside the team—stakeholders or others in the organization—the servant leader takes this responsibility.

Shared leadership is important because it creates effective agile teams, which are empowered and self-organizing, allowing them to make decisions among themselves to set and meet the goals of the team and organization.

Shared leadership is evident when the team solves problems together, determines performance expectations and acceptable interaction patterns, and seeks and evaluates the information that affects how the team functions. Using agile practices and automated tools, the team synchronizes and combines individual team member contributions and makes decisions together regarding the roles team members will take, the plans for product development, and the purpose and frequency of meetings. If these behaviors are lacking, then shared leadership should be improved.

*Peer feedback* is the ability to give accurate feedback to team members based on a common understanding of the team environment. This component is important to correct any errors made by team members, especially in stressful situations. In a development team, errors might include a misunderstanding of user needs, developing code that is not aligned with the design decisions made by the team, or introducing errors in the new code or the existing codebase. Self-correction based on internal feedback is likely to be more efficient, quicker, and more useful than correction from stakeholders outside the team. Moreover, receiving late feedback on errors can generate extra work; a developer might have already moved on to new tasks and will need additional time to recall the context of the error.

A team can use peer feedback to identify its mistakes and offer regular feedback to its members so they can self-correct. Such feedback is typically given in meetings when demonstrating the product, when discussing improvements in a retrospective, or when working closely together, such as when pair programming. Peer feedback can also be given indirectly through, for example, the automated testing of code. A lack of such arenas for feedback indicates a need to improve team effectiveness.

*Redundancy* means that team members have insight into the workload of teammates and use that knowledge to rebalance and reallocate tasks when necessary. For redundancy, knowledge of each team member's responsibilities is needed.

When teams lack redundancy, they cannot readily reallocate their workloads. If workloads rise, some members will work too much, and others not enough. Over time, this can threaten a team's sustainable work pace.

Insufficient progress on tasks or the number of unfinished tasks worked on in parallel may signal a lack of redundancy. A team with good redundancy recognizes its workload distribution problems, shifts task responsibilities, and relieves team members under pressure by reallocating tasks or subtasks to other members.

*Adaptability* refers to a team's ability to adjust its behavior and actions to accommodate changes in the internal or external environment by using the available resources.

Adaptable teams accommodate changes that impact their efforts, and sidestep effects that harm their performance and hinder goal completion. For example, a team unable to adapt to changing requirements could waste time working on the wrong problem. When faced with a technological change, a team must be able to adapt by rapidly learning new technology, or risk becoming less effective.

To evaluate team adaptability, check whether the team takes an interest in the internal or external future events that could affect it. Does the team know the organizational strategy and plans so they can identify potential change? When changes occur, are there arenas to discuss them and plan suitable adjustments?

*Team orientation* refers to the consideration of other team members' behavior when the team interacts and the belief that the goals of the team are more important than the goals of the individual team members. Unlike other teamwork components, team orientation is about attitudes rather than behaviors.

Team orientation is important because it means the team is focused on working together to achieve a single goal, rather than on their individual goals. An individual orientation puts the benefits of teamwork at risk. Without team orientation, heavily interdependent tasks will become harder to complete, and changes to plans may cause the team to struggle. Moreover, rigid roles at the team level can result in team members not engaging in certain tasks and goals; for example, a dedicated test role on a development team might lead to others not taking responsibility for the tasks associated with this role.

As an attitudinal dimension, team orientation might be easier to evaluate based on feedback. A lack of team orientation is evident when team priorities are not considered while choosing work tasks, when there is a lack of interest in offering peer feedback, when assistance is not given to team members with too much workload, or when there is a lack of cohesion.

# How ATEM Can Improve Teamwork

The ATEM can be used in three ways:

- In collocated agile teams, facilitators, agile coaches, and team members can better understand their teamwork by regularly reflecting on how they meet each of the factors in the model, for example during retrospectives. They can use the behavioral markers to evaluate their teamwork and help identify challenges and ways to improve their teamwork.
- Parts of the model may be relevant in contexts beyond single, collocated agile software development teams. What would differ for a very small, distributed, multi-team, or teams doing safety-critical development? A very small team will find it reasonably easy to develop the coordinating mechanisms; a distributed team will need to invest more time to develop an infrastructure that enables good coordinating mechanisms; a team in a multi-team setting would need to ensure that they develop effective coordinating mechanisms with other teams; and teams making safety-critical software would need a special emphasis on peer feedback to reduce errors.
- This model is applicable even if you are on an agile team not doing software development. If your work is knowledge-intensive and similar to software development, or involves many dependencies between work tasks, and if your task priorities change frequently with changes in customer needs or available technology, then ATEM should fit. However, if the context is more traditional, we recommend the original "Big Five" model.[4]

Using ATEM will direct attention to key team effectiveness factors and offer the right thoughts that enable the right actions for agile teams.

## Endnotes

[1]Overeem, Barry. "Characteristics of a Great Scrum Team." scrum.org 15 April 2016. https://www.scrum.org/resources/characteristics-great-scrum-team

[2]Duhigg, Charles. "What Google Learned From its Quest to Build the Perfect Team." *The New York Times Magazine*, 25 February 2016.

# Acknowledgement

We are grateful to the band Franz Ferdinand: our title is a part of their album title "Right words, right thoughts, right action."[b] See the article on ATEM[3] for further acknowledgement regarding research funding. We are also grateful to Anastasiia Tkalich at SINTEF Digital, Marit Larsen at TechnipFMC Norway and Kristin Wulff at Kantega and the Norwegian University of Science and Technology, who provided comments on earlier versions of the manuscript.

# Author biographies

Torgeir Dingsøyr is a software engineering – agile professor at the Department of Computer Science, Norwegian University of Science and Technology. He is an adjunct chief research scientist at the SimulaMet research laboratory. His research focuses on teamwork and learning in software development and development methods for large software projects and programs. He has published in software engineering, information systems, and project management. Contact him at torgeir.dingsoyr@ntnu.no

Diane Strode is a senior lecturer at the School of Information Technology, Whitireia Polytechnic, New Zealand. She is also a research fellow at Open University, United Kingdom. She has experience as a software developer for Mobil Oil Australia. Her research centers on agile software development and coordination, and she publishes in the domains of information systems and software engineering. She has a PhD from Victoria University of Wellington, New Zealand. Contact her at Diane.Strode@whitireia.ac.nz

Yngve Lindsjørn is an associate professor at the Department of Informatics, University of Oslo. He worked for 10 years as a researcher at the Norwegian Computing Center and has 13 years of industry experience as a project manager and a CEO of a software company. From 2009 to 2014, he was the project manager for a research project investigating teamwork in software development. His research includes software development methods and teamwork factors that influence software project success. Contact him at ynglin@ifi.uio.no

---

[b] See https://www.youtube.com/watch?v=RqTsUtQLRFk